\documentclass[prd,nofootinbib, superscriptaddress,preprint]{revtex4}
\usepackage[T1]{fontenc}
\usepackage{amsmath,amssymb}
\usepackage{epsfig}
\usepackage{dcolumn}
\usepackage{graphicx}
\usepackage[usenames,dvipsnames]{color}
\usepackage{slashed}
\usepackage[colorlinks,citecolor=blue]{hyperref}
\usepackage{pdfpages}
\usepackage{float}
\usepackage{adjustbox}
\usepackage[autostyle]{csquotes}
\begin{document}

\title{\boldmath {A new experiment to detect dark matter in sub-MeV range using semiconductor superlattice superstructures (SSS)  }}

	\author{Kalpana Bora}
	\email{kalpana@gauhati.ac.in}
	\affiliation{Department of Physics, Gauhati University, Guwahati-781014, Assam, India}
\begin{abstract}
About  $26\%$ of the matter in our Universe is made up of Dark Matter (DM), which interacts with Standard Model (SM) matter only through gravitational or weak interactions. Many proposals have been made by scientists about the possible candidates of DM - WIMPs, axions, ALPs, black holes etc.  And range of its mass could be extremely broad - from Planck scale to as light as $10^{-22}$ eV. Experiments to detect DM are extremely challenging, as DM does not exhibit appreciable interactions with ordinary matter. May be due to such elusive nature, so far it has not been possible to detect DM, though many experiments are going on worldwide to do so. With about 40 orders of magnitude variation in their mass, it is possible that their gravitational interaction too is very weak, and many creative proposals have been made to detect possible DM candidates, with vast variation of techniques and target materials. In this work, we propose a new experiment to detect sub-MeV range DM particles, using the semiconductor superlattice superstructure (SSS) as the target material. Such materials have band gap of the order of few hundreds of milli eV, and are suitable for detecting sub-MeV range particles scattering off electrons. The photons emitted as a result of excitation of SSS lie in the micrometer range and may be detected via quantum cascade lasers (QCL).\\

keywords : Dark matter, Semiconductor Superlattice Supserstructures (SSS), micrometer  photons, intra-sub-band transitions,  quantum cascade lasers (QCL).\\

\end{abstract}
\maketitle
In spite of the fact that about $26\%$ of the matter in our Universe is made up of dark matter (DM), it has not been possible so far to detect it, may be due to its elusive nature. It is believed to interact with ordinary matter only through gravitational and weak interactions. Various proposals have been made by researchers for the possible dark matter candidates, that range from axions/axion like particles (ALPs) to black holes. Accordingly, mass of dark matter candidates may range from $\sim 10^{-22}$ eV to the Planck scale. Many experiments have been performed/planned by scientists worldwide,  that include both direct detection (DD) and indirect detection experiments and production of DM candidates in accelerators. In DD experiments, the DM particle scatters off the nucleus or electrons in the atom. For a recent review on DM searches, please refer to \cite{Billard2021}-\cite{Heros2020} (and references therein).\\
\\
However, there have been stringent exclusion limits and null results from most of current DM detection experiments in the TeV range DM candidates \cite{Kahn2021}. Thermal DM can have mass range (keV-100 TeV),  with WIMPs (weakly interacting massive particles, produced thermally via freeze-out) lying in range (1 GeV-100 TeV). A DD experiment looks for kinetic energy deposited in nucleus of the target in DM scattering off the target nuclei. If $\rho_\Phi$ is the density of DM, $m_\Phi$ is its mass and $n_\Phi$ its number density, then, since the DM penetrates through our galaxy, then \cite{Kahn2021},

\begin{equation} 
 \rho_\Phi=m_\Phi\times n_\Phi \sim 0.3-0.5 \hspace{0.1 cm} GeV/cm^3,\hspace{0.1 cm}v_\Phi\sim10^{-3}\textrm{.}
\label{Eq1}
\end{equation}\\
The DM-nucleus scattering cross-section $\sigma_{\Phi N}= A^2\alpha_W^2/m_\Phi^2$, where $A$ is the atomic mass number of the target material, $\alpha_W\sim 0.03$ is the coupling constant. Velocity of bound DM in our galaxy is limited by the galactic escape velocity $\sim 10^{-3}c$. Hence, for a heavy nucleus of $A=100$, and for a DM mass $m_\Phi=100$ GeV,  scattering rate per nucleus $R_\Phi$ turns out to be,

\begin{equation} 
R_\Phi=n_\Phi\sigma_{\Phi N}v_\Phi \textrm{,}
\label{Eq2}
\end{equation}\\
and can be computed to be $\sim 3\times10^{-26}/s$. Therefore, huge detectors are needed to have significant number of events per year. As the DM is non-relativistic, for DM mass $m_\Phi=100$ GeV, its momentum $p_\Phi v_\Phi$ and kinetic energy $m_\Phi v_\Phi^2/2$ are such that this DM-nucleus scattering is elastic. The WIMP DM below GeV is unaccessible to current experimental searches, but presents opportunities for new experiments. The detection of sub-GeV DM is very challenging because of insufficient  signal due to less recoil energy. For DM mass $\sim$ few MeV, the nuclear recoil energy lies in the range $\sim$ eV, that is of the order of band gaps of semiconductors. Also, as DM mass $m_\Phi$ decreases, its number density increases, and hence smaller size detectors can also provide sufficient event rate for DD experiments. WIMPs with very low mass, of the order of $\leq$ MeV cannot produce sufficient nuclear recoil (NR) energy, which lies below the threshold of most experiments,  and hence coupling of DM to electrons (electron recoil, ER) is used to detect such light DM particles. Condensed matter systems are also used to detect light DM \cite{Kahn2021}, as bridging atoms together lowers the excitation energy, and condensed matter have large densities.\\
\\
Above discussion forms the motivation for this work, in which we have proposed a new idea for detecting sub-MeV range DM, using ER in SSS materials. The band gap of such materials is few hundreds of meV (milli eV), which matches with electronic recoil energy for scattering by sub-MeV DM. When the SSS target after excitation by DM comes to ground state, the photons emitted are in micrometer range, which can be detected by quantum cascade lasers.\\

The nuclear recoil energy in DM-nucleus scattering (elastic) can be expressed as

\begin{equation} 
E_{NR} =\frac{q^2}{2m_N} \leq \frac{2\mu_{\Phi N}^2v_\Phi^2}{m_N} \textrm{,}
\label{Eq3}
\end{equation}\\
where $\mu_{\Phi N}$ is the reduced mass of the DM-nucleus system, and $v_\Phi$ is the DM velocity which is taken as sum of galactic escape velocity and earth's velocity to estimate the maximum nuclear recoil energy. This recoil energy becomes insufficient and falls below threshold of most current generation experiments to detect lower mass (sub-MeV) DM. For such DM, energy is transferred more efficiently to an electron \cite{Essig2012}. The electron recoil energy is

\begin{equation} 
E_{ER} \leq \frac{\mu_{\Phi}v_\Phi^2}{2} \leq 3.23\hspace{0.1 cm} eV ( \frac{m_\Phi}{MeV})\textrm{,}
\label{Eq4}
\end{equation}\\
and for a DM of mass 0.2 MeV, the electronic recoil energy is 255 meV (shown in Table 1) if DM velocity is taken as $2.54\times10^{-3}c$ (as sum of galactic escape velocity and speed of earth, which turns out to be $\sim764$ km/sec \cite{Battaglieri2017}). This recoil energy can promote bound electrons from valence bands to higher bands if the band gap is of the order of $\sim$ 255 meV, and  when they return back to their ground state, micrometer (terra hertz) photons are produced,

\begin{equation} 
\lambda=hc/E = 12.3\hspace{0.1cm}\textrm{micrometer (25 THz), for E=100 meV.}
\label{Eq5}
\end{equation}\\
Next, a mechanism is needed to detect such micrometer photons, because in the scattering experiment, detection of such photons will act as the signal (signature if a DM has reached the detector). From this discussion, we observe that for DM of typical mass $\sim$ 0.2 MeV, the recoil energy for DM-electron scattering is of the order of 250 meV, and hence we need a target with such band gaps. In this work, we propose a novel experiment, where SSS can be used for the purpose, as they have the desired band gap of hundreds of mill eV. \\

\begin{table*}
\centering
\begin{adjustbox}{width=10 cm}
\begin{tabular}{|c|c|c|}
\hline 
DM Mass & Scattering by Xenon nucleus & Scattering by electron  \\ 
\hline 
0.2 MeV & 0.00196 meV & 255 meV\\ 
\hline 
0.5 MeV & 0.00123 meV & 806 meV\\ 
\hline 

\end{tabular}
\end{adjustbox}
\caption{Recoil energy for DM scattering by Xenon and electron}
\label{tab:Table 1}
\end{table*}

We give here an overview about SSS for the purpose of completeness of the work \cite{Smith1990, Parker2005}. Superlattice is a periodic structure of layers of two or more materials (say AlAs and GaAs layers, also GaInAs SSS), of similar lattice constant and different band gaps. Typically, width of layers is orders of magnitude larger than the lattice constant. There are in general two methods of forming superlattice structure - periodic variation of donor or acceptor impurities, or periodic variation of alloy composition introduced during crystal growth. In the first method, due to larger thermal diffusion of impurities, it is hard to maintain periodic nature of the potential, and hence the second method is more popular. It should be noted that superlattice is not same as multiple quantum well, because though they are very similar in their structure, the barrier width of SSS is small enough (than quantum well), such that quantum wells are coupled with each other. On the other hand, quantum wells in multiple quantum well structures are highly localised. The solution of Schroedinger's equation using Kronig-Penny model gives discrete energy bands separated by forbidden bands, just as in usual crystal structure energy band. Difference is - here, the band gap can be engineered by controlling the widh of alternating materials, and also by choosing the materials of alternating layers. Because the thickness of alternating layers is orders of magnitude larger than the lattice constant, the energy bands form in the first Brillouin zone of the well, and first Brillouin zone is subdivided into many mini-bands. These mini-bands are the main characteristic of superlattice structures. The potential barrier of few nanometer can create a band gap of few hundred milli eV in typical SSS (like GaInAs). This band gap corresponds to micrometer region photons of the spectrum, which will be emitted if a sub-MeV DM candidate is scattered off such SSS target.\\

\begin{figure*}
  \centering
\includegraphics[scale=0.6]{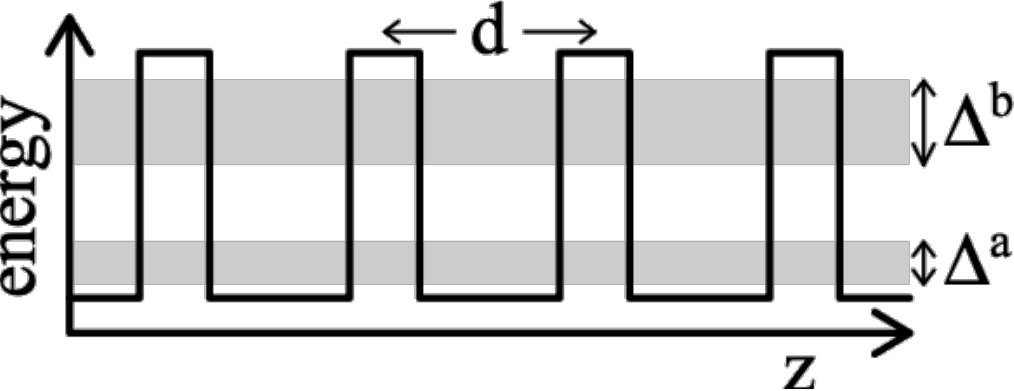}
\caption{\ Spatial variation of conduction band edge E(z), with mini bands $\nu$=a,b (shaded areas) for semiconductor superlattice \cite{Wacker2002}}
\end{figure*}

\begin{figure*}
  \centering
\includegraphics[scale=0.7]{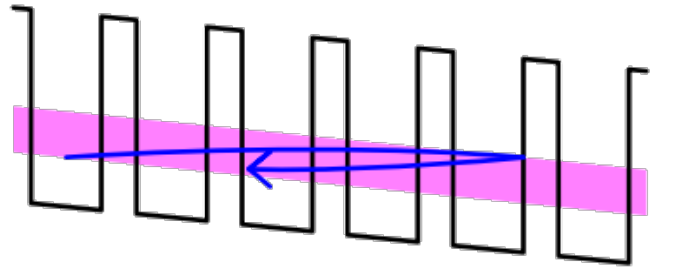}
\caption{\ Mini band conduction mechanism for superlattice transport  \cite{Wacker2002}}
\end{figure*}

\begin{figure*}
  \centering
\includegraphics[scale=0.4]{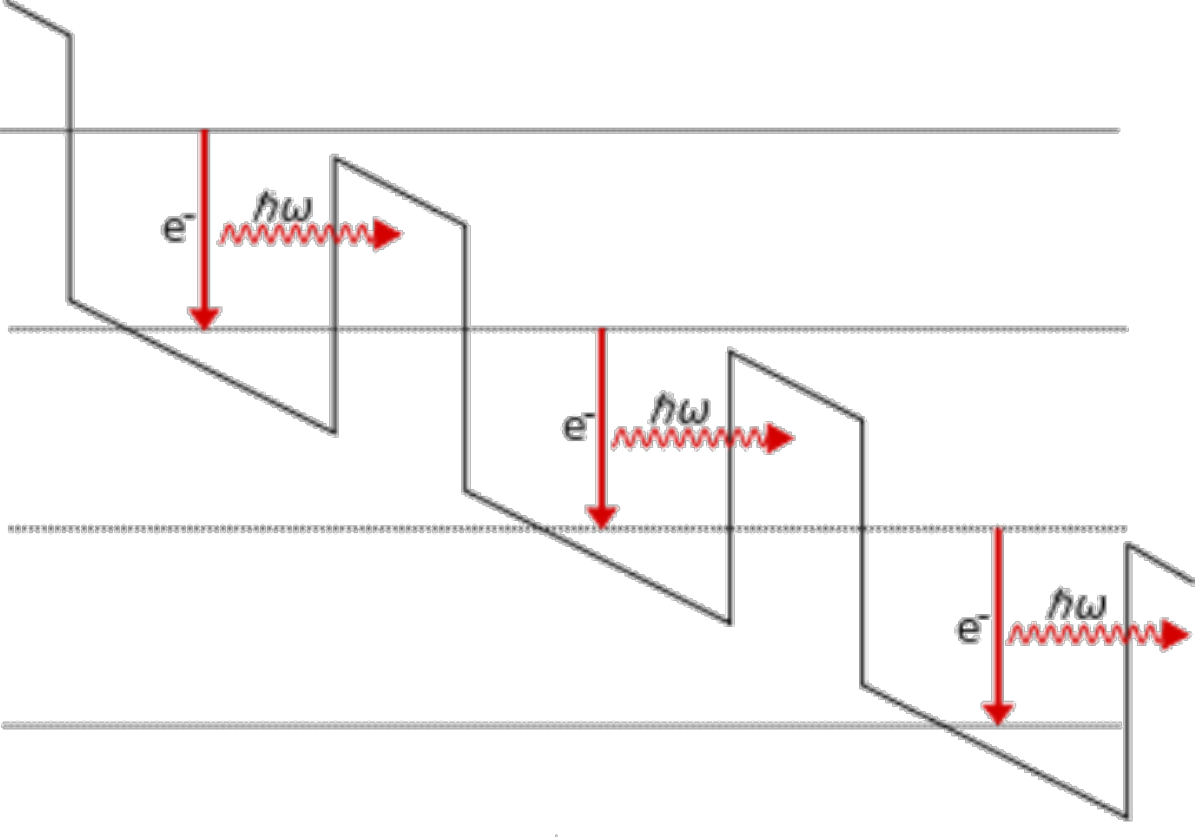}
\caption{ Intra sub-band transitions and laser mechanism in QCL}
\end{figure*}
\vspace{0.1 cm}

Next challenge is to detect such micrometer photons, and  QCL can be used for the purpose. SSS are used in the making of QCL. In QCL, intra-sub-band transitions of SSS are used, in place of inter-band transitions used in traditional lasers. In the intra-sub-band transitions, carriers do not recombine and disappear, rather they reside in the quantum well and can still be transported. Electrons stream down a potential staircase and emit photons at steps. These steps consist of quantum wells,  in which control of tunnelling yields population inversion between discrete conduction band excited states \cite{Faist1994}. Direct evidence of laser action at micrometer wavelength, with peak power in pulsed operation, is exhibited by a strong narrowing of emission spectrum above threshold - and this acts as the signal that a micrometer photon has been detected. Further details about how this experiment can be realised, is beyond the scope of this work, and can be worked out by experimentalists of the field.\\

\begin{figure*}
  \centering
\includegraphics[scale= 1.2 ]{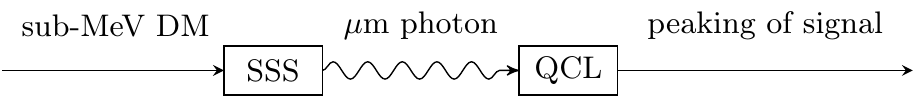}
\caption{\ Schematic diagram of the proposed experiment}
\end{figure*}

To summarise, we have proposed a new experiment to detect sub-MeV range DM, using SSS, where the micrometer range photon as signal can be detected using QCL. The band gap of SSS materials can be engineered by adjusting width of alternating materials and with change of such materials, such that the DM in a mass range can be detected (equivalent to detection of photons with a range of wavelengths). We have just proposed the idea for a futuristic  experiment, and further details/challenges in designing etc. lie beyond the scope of this paper and can be worked out by the expert experimentalists in the field. Hope this proposal will open new avenues in the search and detection of DM.

\section*{\textbf{Acknowledgements}}

The author would like to thank the support by RUSA and FIST grants of Govt. of India for upgrading the computer lab of the department where part of this work was done.


\end{document}